# Automatic Web Security Unit Testing: XSS Vulnerability Detection

**Mahmoud Mohammadi**
University of North Carolina at Charlotte
Charlotte, NC, USA
mmoham12@uncc.edu

**Bill Chu**
University of North Carolina at Charlotte
Charlotte, NC, USA,
billchu@uncc.edu

**Heather Richter Lipford**
University of North Carolina at Charlotte
Charlotte, NC, USA
richter@uncc.edu

**Emerson Murphy-Hill**
NC State University
Raleigh, NC, USA
emerson@csc.ncsu.du

## ABSTRACT
Integrating security testing into the workflow of software developers not only can save resources for separate security testing but also reduce the cost of fixing security vulnerabilities by detecting them early in the development cycle. We present an automatic testing approach to detect a common type of Cross Site Scripting (XSS) vulnerability caused by improper encoding of untrusted data. We automatically extract encoding functions used in a web application to sanitize untrusted inputs and then evaluate their effectiveness by automatically generating XSS attack strings. Our evaluations show that this technique can detect 0-day XSS vulnerabilities that cannot be found by static analysis tools. We will also show that our approach can efficiently cover a common type of XSS vulnerability. This approach can be generalized to test for input validation against other types injections such as command line injection.

## Categories and Subject Descriptors
D.2.5 [**Software Engineering**]: Testing and debugging – *Code inspections and walk-throughs.* D.4.6 [**Operating Systems**]: Security and Protection – *Verification.*

## 1. INTRODUCTION
In web applications data from untrusted sources, such as user provided profile and comments, are often displayed on web pages. An attacker can exploit such applications by providing information containing malicious JavaScript programs that can be executed to cause harm, such as stealing innocent user's login credentials. To prevent such Cross Site Scripting (XSS) attacks, one of the most common security attacks today, web applications should sanitize untrusted data using output encoding functions before displaying them on web pages. Static analysis techniques[14] are widely used to check whether a web application uses encoding functions to sanitize untrusted data. However, static analysis cannot verify whether the encoding functions are used correctly. Testing must be performed to ensure that the encoding function is used effectively to prevent XSS attacks. In this paper we consider web applications written in Java and JSP. Our approach can be extended to web applications written in other languages as well.

To successfully prevent XSS attacks, encoding must match the context in which untrusted data appears. The following contexts have been identified for a typical web application: HTML body, HTML attribute, CSS, URL, and JavaScript[15]. Well-tested encoders have been written for each of these contexts. A common programming error is that one chooses a wrong encoder for a given application context. Consider the fragment of JSP program in Figure 1. Native Java code is enclosed in <% %>.

```
1) <% String pid = (String)request.getParameter("pid");  %>
2) <% String x = (String) request.getParameter("addr");%>
3) <a href="javascript:void(0)"
   onclick="action(<%=escapeHtml( pid ) %>');" >  mylink
   </a>
4) <p> <%=escapeHtml(x) %>
```

**Figure 1. Motivation Example**

This example has two user provided input: pid and x. Variable pid is used as part of rendering an HTML anchor element on line 3, and x is displayed in the HTML body on line 4. A maliciously supplied input for x might be *<script> attack(); </script>*. If the encoding function, *escapeHtml()*, were not applied, this would cause the execution of the JavaScript function *attack()*. Encoding function *escapeHtml()* would replace < and > characters with *<* and *>* respectively and transform the malicious input into *<script> attack(); </script>* thus preventing *attack();* from being interpreted as a JavaScript program by the browser.

However, the same encoding function does not work for the case on line 3. A malicious input for pid might be *'a'); attack(); //* and it will pass *escapeHtml()* unchanged. The rendered anchor element would be *<a href= "javascript:void(0)" onclick="action('a');attack();//"> mylink </a>* invoking JavaScript function *attack()* when the link is clicked. The reason for this vulnerability is that the wrong encoder function is used for this context. The correct encoder, a JavaScript encoder, would replace the single quote character with *\'* thus preventing this attack. One Google sponsored study showed that 30% of encoding function usage is incorrect [11] leading to serious vulnerabilities for web applications. There are also cases where more than one encoding function must be used (an untrusted input occurs in two contexts simultaneously, e.g. JavaScript context and URL context). The order of applying encoders is sometimes important.

To address this problem, a best practice-programming guide has been published by the OWASP foundation [9] to inform developers to match encoders to web contexts. However, there is no systematic way to test whether encoder functions are used correctly. Penetration testing, which often happens late in the software development cycle, is relied upon to test for XSS vulnerabilities.

Other efforts have looked at using type inference to detect the context of an untrusted variable so the correct encoding function can be and automatically applied [11]. To aid type inference, such efforts all work with template languages[11], such as Closure Templates or HandleBars, that have stronger type systems than W3C languages. There are at least three limitations of this approach. First, a template language covers a subset of available web technologies. Second, there are many web applications that do not use template languages at all. Third, type inference is not fully successful. For example a research team from Yahoo! [7] found that they could identify the correct context in about 90.9% of applications written in HandleBars. A Google sponsored research[11] also showed that type inference is not possible for some cases for applications written in Closure Templates, which they proposed a runtime auto-sanitization mechanism that may incur significant (9%) run time performance penalty. Other researchers [2,3,12] used similar combinatorial pattern-based attack generation mechanisms to test XSS vulnerabilities. By focusing on unit testing of encoding functions, our approach can not only discover vulnerabilities much earlier in the development cycle but also provide precise feedback to developers to fix the vulnerability. We have a clearly defined vulnerability model which provides a basis for more formal analysis of test coverage. Related works relied on analysis of test output to evaluate tests, our test evaluation is based on execution behavior which can discover subtle vulnerabilities originated from internal browsers decoding function mentioned in [8,15]. Our approach works with all existing web languages. Integrating security testing into the workflow of software developers not only can save efforts on separate security testing but also reduces the cost of fixing security vulnerabilities by detecting them early in the development cycle.

## 2. APPROACH
This section provides a detailed description of our approach. We first more precisely define the type of vulnerability we are aiming to detect via automatic testing. We then describe three key steps of our approach: extracting unit tests from application source, evaluating unit tests, and generation of attack strings.

### 2.1 Vulnerability Model
We assume the web application follows best secure programming practice[9] by using correctly implemented encoding functions. An encoder function is designed for a given web context, such as HTML body and replaces characters that are unsafe in that context. An unsafe character is a special character that may be interpreted by the browser as a part of the markup language. For example character < is unsafe in an HTML body context because it can start a new HTML element. Character < is replaced by *<* in an HTML body by HTML encoder. Character single quote is considered safe in HTML body context and thus not changed. However, single quote is unsafe for the JavaScript context because it can be used to launch attacks.

We define a variable **x** as *tainted*, if it is assigned a value from untrusted inputs such as *request.getParameter()*. Common sources of untrusted input for web applications are well known to commercial static analysis tools. In this paper, we define a *security sink* as a Java statement generating output containing a tainted variable in JSP. These Java statements in JSP can be easily identified as they are marked by *<%= ... %>* tags.

*Encoder f is **safe** for context C means for **all** possible input x such that x contains unsafe characters in context C, f(x) doesn't contain unsafe characters in context C (this implies f(x)!=x).* The type of vulnerability we seek to find using unit tests involves an attack string that can successfully pass through an encoder unchanged.

*Encoder f is **vulnerable** for a context C means **there is** an input x such that f(x)=x ( from the browser's view), and x contains a JavaScript program that can be successfully executed in C .*

We put the caveat "from the browser's view" because browsers may treat certain characters as equivalent in specific contexts. For example ' is equivalent to single quote character in the HTML body context. Given this vulnerability model, our goal is to develop an attack generator such that if the encoder function is vulnerable, a successful attack string can be found. We envision this to be implemented as an IDE plug-in to automatically build security "unit test cases" based on encoding function usages and then evaluate them by applying generated attack strings.

### 2.2 Unit Test Extraction
We extract security unit test cases, or simply referred to as a *unit tests* automatically based on application source code using data and control flow analyses. A unit test is centered on the encoding of a tainted variable. It should contain sufficient web application context so that we can test whether the encoder is vulnerable for this context. Therefore, a unit test is a set of statements that starts with the introduction of a tainted variable and ends with putting the tainted variable in a security sink. Again, we define a sink as an output generating Java statement containing *f(e)* where *f* is an encoding function and *e* is an expression containing a tainted variable. Expression *e* always evaluates to a string. For this paper, we assume each sink contains one tainted variable and one sink in a unit test. Our approach, however, can be generalized in the future to include multiple variables.

```
1.   <% String pid = (String)request.getParameter("pid");
2.   if( pid.startswith('2015') )
3.       pid =session.getAttribute("group") + pid;
4.   else
5.       pid =session.getAttribute("OU") + pid; %>
6.   <table><tr><td>
7.    Project ID=  <%=escapeHtml(pid ) %> </p>
8.   <% for ( int c=0; c < Tasks(pid).length(); c++ ) { %>
9.   <a href="javascript:void(0)" onclick="action( c , '
     <%=escapeHtml( pid ) %>');" > Details  </a>
10.  <% } %> </td></tr></table>
```

**Figure 2. Source Code Under Test**

We illustrate the unit test extraction process through the example in Figure 2. The goal is to build an executable unit that only contains application logic surrounding the encoding function. In this example there are two security sinks (lines 7 and 9), each corresponds to a different unit test. Figure 3 is the unit test based on the security sink of line 9 in Figure 2.

Variable *pid* is tainted because it is originated from user input (*request.getParameter*). The test case includes the security sink as well as any W3C language elements (CSS, HTML, JavaScript etc.) necessary to execute the security sink. In Figure 2 these language elements include the anchor tag (starting with {<a href=…} on line 7), and if statement from line 2 to line 5 of Figure 2.

```
UnitTest1.JSP
1.   <html><body> <script> function Fn(x) { return; } </script>
2.   <% String pid = "Constant" + request.getParameter("atk");
3.   if( pid.startswith('2015') )
4.       pid  ="" + pid;
5.   else
6.       pid  ="" + pid;  %>
7.   <a href="javascript:void(0)" onclick="Fn(' <%=escapeHtml(
     pid ) %>');" id="tagid">  </a>
8.   <% } %> </body></html>
```

**Figure 3. Extracted Unit Test**

The *if* statement is included because it is related to variable *pid*. Any variables (e.g. *session.getAttribute("group")* on lines 3 and 5 in Figure 2) are replaced with null string giving the attacker the best chance to succeed.

Special attention is given to cases where the security sink is a parameter to a JavaScript function, as is the case on line 9 in Figure 2, where an application specific function *action(_,_)* with two parameters is used. It is sufficient to test if a successful attack can be found against a generic JavaScript function with only one parameter. If such an attack is successful then the wrong encoder is being used and an attack for the original function can easily be constructed. Therefore, we replace any JavaScript function containing the security sink (e.g. function *action(_,_)* in Figure 2) with a predefined function having one parameter returning a null string value, as illustrated on line 7 in Figure 3. This predefined function *Fn()* is defined on line 1 of the Figure 3.

The for-loop on line 8 is not included because it does not impact any values in executing the security sink.

## 2.3 Attack Evaluation

The goal of attack evaluation is to assess whether the extracted unit test is vulnerable to any of the generated attack strings. We utilize JWebUnit features to execute the test web page and check the result of invoked web page to verify whether each attack string generated is successful. Our attack payload changes the title of the web page if it is successfully executed. So if a change in title is detected, then the attack string is successful.

We use JWebUnit because some XSS vulnerabilities are only revealed when the attack script is executed in a real browser. Also we must simulate user interface behaviors (such as mouse clicks) because some attacks are only successful upon user-generated events. These testing can also be done using other testing libraries such as PhantomJS or CasperJS if they can be fully integrated in a testing framework such as JUnit to be used in a IDE like Eclipse.

A unit test driver is created for each unit case. Figure 4 shows test driver for the test code in Figure 3. After initializing an instance of WebTester (subclass of JWebUnit) on line 1, line 2 sets the base url of the application under test. Each iteration of the loop on line 3 takes one generated *attackString* and invokes the unit test page (UnitTest1.jsp). Line 5 is included only when the test code includes an event, "onclick" as in this example. The test driver needs to simulate the user action of clicking the link to see if any attack is triggered by the clicking event. The next line clicks the hyperlink in unit test (line 7 in Figure 3). If page's title equals to "ATTACK" then the encoder function is vulnerable for this application context.

```
Java Code to Invoking UnitTest1.jsp:
1.    WebTester tester =  new WebTester ();
2.    tester.setBaseUrl("Address Of Main App");
3.    for( String attackString: generatedAttackScripts) {
4.      tester.beginAt("UnitTest1.JSP?atk="+ attackString);
5.      tester.clickLink("tagid");
6.      // Attack vector changes the page's title to "ATTACK". It
   is required to check whether it is changed or not.
7.      tester.assertTitleNotEquals("ATTACK");
8.    }
```

**Figure 4. Attack Evaluation**

## 3. ATTACK GENERATION

The goal of attack generation is to generate attack strings with JavaScript programs as input to the security unit tests. We define the attack string as composed of *pre-escaping characters, attack payload,* and *post-escaping characters* substrings. Figure 5 shows an example of this pattern.

| Pre | Attack Payload | Post |
|---|---|---|
| '); | attack(); | // |

Original Source:

*<input type='button' onclick="fn('<%= escapeHtml(x) %>'); " />*

After Attack:

 *<input type='button' onclick=" fn(' '); attack(1); // '); " />*

**Figure 5. Basic Attack Vector Pattern**

The attack payload can be any valid JavaScript statement and pre and post strings are string literals required to manipulate the browser to correctly parse and then execute the attack payload as a valid JavaScript statement. Pre and Post escaping strings are thus key elements to generating a successful attack string. A web page consists of HTML elements, which can be described by W3C published XML grammar[13] shown in Figure 6.

| element | ::= STag **content** ETag |
|---|---|
| STag | ::= '<' Name (Attribute)* '>' |
| Attribute | ::= Name Eq **AttValue** |
| **AttValue** | ::= '"' ([^<&"] \| Ref)* '"'  \|  "'" ([^<&'] \| Ref)* "'" |
| **Content** | ::= CharData? ((element \| Ref \| CDSect \| PI \| Comment) CharData?)* |
| ETag | ::= '</' Name '>' |

**Figure 6. XML elements grammar**

An attack string can appear in one of three contexts: as part of **attValue**, part of **content**, or part of a JavaScript program. Attacks cannot be part of the ETag because it does not allow any attributes. We refer to these three options as attribute values, tag bodies and JavaScript. Attack strings must be generated for all three contexts. Our approach to generate attack strings is based on a finite state machine. The state machine is constructed based on an empirical model for a generic browser developed by[15] as well as W3C published language specifications[13]. A browser has an interpreter for each of the languages it processes: HTML, CSS, JavaScript, etc. Upon reading certain tokens from the input stream, a given interpreter may call upon another interpreter, a process referred to as context switching. For example an HTML interpreter, upon reading the tokens *<script>* invokes the

JavaScript interpreter to interpret a JavaScript program, or switching from an HTML body context to a JavaScript context. An attack string in either the attribute context or tag body context thus uses a sequence of pre-escaping characters to manipulate the browser to invoke the JavaScript interpreter. If the attack string is already in a JavaScript context, it uses a sequence of pre-escaping characters to insert the attack payload as an additional JavaScript statement. Figure 7 illustrates context switching using an example. Suppose an interpreter has already consumed characters $t_1t_2t_3$ and currently is in context C1. The goal of attack string, $t_4t_5…t_n$, is to change the context using the single quote character to indicate the end of an attribute value, followed by inserting a tag event key word (onclick) which switches to a JavaScript context. After that switch, a JavaScript program can be inserted.

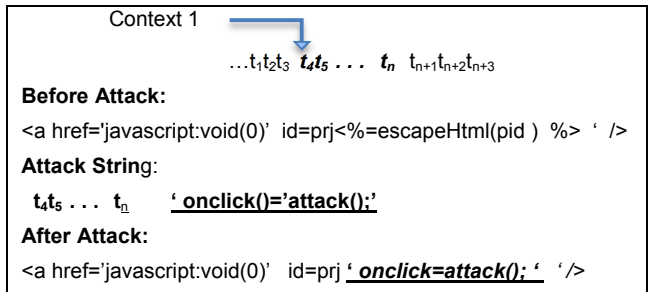

**Figure 7. Context Changing**

In the state machine for attack generation, states correspond to states of an interpreter. Starting states corresponds to the three contexts where attack scripts can be injected: attribute value, tag body, and JavaScript. The state machine is built with the objective of enumeration all possible transitions from a start state to a context where a JavaScript program can be executed. Transitions are based on grammars of web languages (HTML, CSS, URL, JavaScript). A token list is associated with each transition, representing possible input tokens that can take the interpreter from one state to another state along the path to the objective. Final states, indicated by small bold circles, represent completion of an attack string.

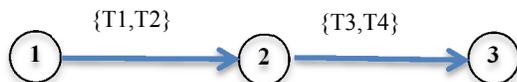

**Figure 8. Sample State Machine**

An attack script is generated by traversing the state machine while appending tokens associate with transition link to the output string. The number of tokens in transition lists determines number of different combinations or attack vectors. For example, having different tokens listed on links between states, by traversing from S1 to S2 to S3 in sample state machine shown in Figure 8 we would have 4 different output strings: T1T3, T1T4, T2T3 and T2T4.

Figure 9 illustrates the state machine for attack generation. We explain the state machine based on the three possible starting points of a successful attack.

## 3.1 Attribute Value Context

This part of the state machine considers the situation where an attack string is injected in the middle of a tag definition (e.g. the definition of an <input> tag), or the tag attribute context. Table 1 lists states for the attribute value context. Attribute value context is located with the *STag* definition of the XML grammar in Figure 6.

Starting from state S1, we have two possibilities to change the context to JavaScript. The first option (1.1) is when the attack string is located as part of a predefined special attribute (one of: style, href, src). In this case any tokens labeled by **Ctx.Keywords** in Table 2 can take the interpreter into a JavaScript context and then final state. For clarity, each token is enclosed by {}. String {%V%} in **Table 2** represents a place where an attack payload can be inserted. This path generates three possible attack strings: { *javascript:attack();* } , { *url('javascript:attak();')* } and {expression('attack()')} with the last one targeting IE7 and earlier.

**Table 1. Attribute Value States**

| State | Description |
| --- | --- |
| S1 | Starting point of state machine for injection into attributes' value |
| S2 | The Tag's valued finished and can add a new attribute or closing the tag |
| S3 | A special attribute (e.g. CSS, URI enabled) added and ready to switch to JavaScript context |
| S4 | Context changed to JavaScript |
| S5 | Starting point of state machine for injection points in tag's body. |

**Table 2. Attribute Value & Tag Content: Labels**

| Transition | Tokens |
| --- | --- |
| Att.Marker | {'},{"},{' "},{" '} |
| End.Tag | { >} ,{ />} |
| Tag.Starter | {<a }, {<img } |
| Att.Starter | { atb=} , { atb='} , atb="} |
| Event | { onclick='%V%' } |
| Ctx.Keywords | {url('javascript:%V%')}, {javascript:} , {expression('%V%')} |
| Spec.Att | { src=},{ style=}, {href=} |
| Start.Script | {<script>%V%</script>},{</script><script>%V%</script>} ,{</title><script>%V%</script>} ,{</textarea><script>%V%</script>} |

The second possibility (1.2) in state S1 is to close the current tag attribute to start a new attribute accepting JavaScript programs. Tokens signaling the end of tag attributed are labeled **Att.Marker** in Table 2. This takes the state machine to state S2. According to the *STag* grammar rule, at this point one can have three options: first (2.1) closing the tag using '>' (**End.Tag**) and be in the tag body context represented by state S5 or second (2.2) adding an event attribute, and the third option (2.3) adding special attributes. Selecting the second option (2.2) in state S2 can be done by using tokens labeled *Event* in link between S2 and S4.

We only generate one representative event, 'onclick', as all other events work in the same way as far as running a JavaScript program is concerned. Reaching state S4 means that we are in a JavaScript context and can place our attack payload indicated by %V%.

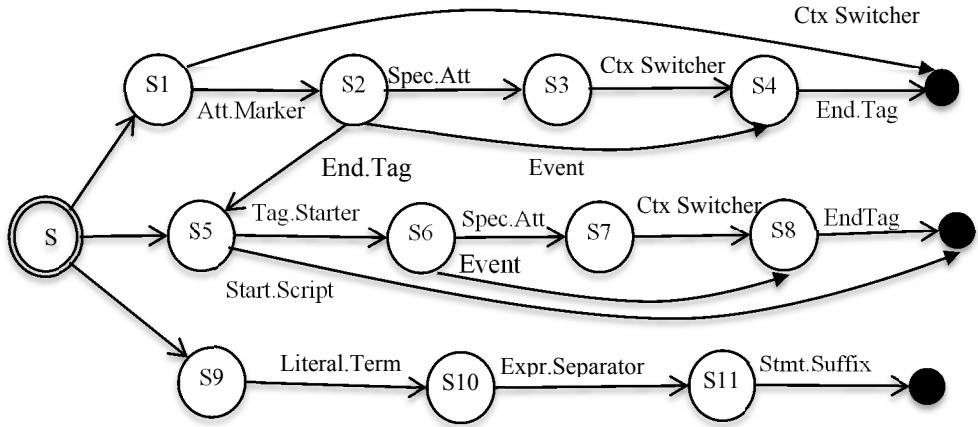

**Figure 9. Attack generation State Machine**

After state S4 the tag attribute definition is complete and we can close the tag with tokens in the ***End.Tag*** label between state S4 and the final state. In summary, this path {S1 + S2 + S4 +Final State} generates attack scripts by appending tokens of the { *Att.Marker + Event + End.Tag* }, or **'onclick=attak(); >** as one of the possibilities.

Option (2.3) in state S2 is adding a special attribute. In this option ***Special.Att*** label takes interpreter to state S3 to add one of the special attributes (src, href). These attributes have the built in capability to switch to JavaScript. Once we are in state S3 we use any tokens labeled by ***Ctx.Keywords*** that can take the interpreter into a JavaScript context in state S4. In summary, the path {S1 + S2 + S3 + S4 +Final State} generates attack scripts by appending tokens of { *Att.Marker + Spec.Att + Ctx.Keywords + End.Tag* }, with **'** *href= javascript:attack(); >* as an example.

### 3.2 Tag Content Context

This part of the state machine considers the situation where an attack string is injected in the middle of a tag body (e.g. <td>, <title>, or <style>), corresponding to ***content*** rule in XML grammar definition of 6. Table 3 summarizes all the states related to this scenario.

**Table 3. Tag Content States**

| State | Description |
|---|---|
| S5 | Starting point of state machine for injection into tag's body. |
| S6 | The Tag's valued finished and can add a new attribute or closing the tag |
| S7 | A special attribute (e.g. CSS, URI enabled) added and ready to switch to JavaScript context |
| S8 | Context changed to JavaScript |

From start state S5, two options are available to switch to execute a JavaScript program: (5.1) using *<script>* tag to switch to JavaScript context or (5.2) creating a new tag with a JavaScript program as part of its attribute value.

Option (5.1) is represented by the transition between S5 and the final state with tokens labeled **Start.Script** in Table 2. This transition considers two possible situations. The first case is the attack string is in a tag body where a JavaScript program can be inserted using the *<script>…</script>* pair tags. The second situation is the attack string is in a tag body that does not accept the *<script>…</script>* pair tags. Based on relevant W3C specifications, the latter case can occur if the attack string is in these tag bodies: *<script>* (nested scripts are not allowed), *<title>*, and *<textarea>*. These tags must be closed in order to invoke the attack payload. The following attack string *</title><script> Attack(); </script>* is an example of this option.

Option (5.2) means that we start a new tag with Tag.Starer label( Table 2) and transition to S6. Several tags can take JavaScript programs as an attribute value; most of them are syntactically isomorphic so we do not need to consider all of them. We choose the achor (<a) and image (<img) tags because they have the most attribute options (e.g *href*, src*)*. Other possible attacks can be syntactically mapped to attack strings starting with these tags.

S6 is a state where a new tag has been started. This state is similar to state S2 discussed in the previous section. Instead of transitioning to S2, we create S6 because W3C does not allow nested tags. At this point we have two options to switch to JavaScript context one option (6.1) adding events and option (6.2) adding special attributes.

Option (6.1) leads to this path: {S5 + S6 + S8 +Final State}. This path generates attack scripts by appending tokens of the {Tag.Startter + *Event + End.Tag* }. For example: **** is a generated attack string.

Option (6.2) leads to path {S5 + S6 + S7 + S8 +Final State} which generates attack scripts by appending tokens of { *Tag.Startter + Spec.Att + Ctx.Keywords + End.Tag* }, with **<a href= javascript:attack(); >** as an example**.**

### 3.3 JavaScript Context

This part of the state machine considers the situation where an attack string is injected into the middle of a JavaScript Program. An attacker will have to insert pre-escaping characters to end the current JavaScript statement and insert an attack statement. Tables 4 and 5 summarize states and transition labels for this attack category.

In this paper we considered two common types of attacks in the JavaScript context: the attack string is part of an expression on the right hand side of an assignment statement, or the attack string is part of an expression that is parameter to a function call. Our approach can be extended to cover all possible attacks in the JavaScript context.

**Table 4. JavaScript States**

| State | Description |
|---|---|
| S9 | Attack string injected into JavaScript context |
| S10 | Current expression containing injection point joined to a new attack payload expression |
| S11 | Attack payload added and the remaining characters( if any) should be commented out |

**Table 5. JavaScript Labels**

| Label | Tokens |
|---|---|
| Literal.Term | { ' } , { " } |
| Exp.Seprator | { + (%V%)} ,{ ; %V% } , { ); %V%)} |
| Stmt.Suffix | Null, { // } , { ); // } |

In either case the attack string is part of a string literal that is enclosed either by a pair of double quotes, or a pair of single quotes, as defined by JavaScript language specifications [4]:

```
StringLiteral ::= "StringChars"|'StringChars '
```

To insert the attack payload, the pre-escaping must first terminate the string literal it is part of. This can be accomplished by adding either a single quote or a double quote to end the string literal: transitioning from S9 to S10 using literal terminator tokens labeled by **Literal.Term** in **Table 5**. After closing the string literals we are in state S10 and ready to add the attack payload. Two cases are possible: (10.1) attack string is part of an existing expression, or (10.2) the attack string needs to start a new statement.

Tokens to accomplish this goal are listed under the **Exp.Separator** label of **Table 5** leading us from state S10 to S11. For 10.1, one can use an operator, e.g. the string concatenation operator +. For option 10.2, two cases are possible. If the attack string is part of an assignment statement, a semicolon can terminate the assignment statement {;}. If the attack string is part of a function call, it can be terminated by {);}. Once we are in state S11 we can add the attack payload as indicated by %V% placeholders in **Table 5**.

### 3.4 Discussions on Attack Generation

We have built a state machine based model to generate XSS attack strings, based on published W3C grammars for JavaScript and related markup languages. Our goal is to generate a set of attack strings that can "cover" all possible successful attacks in the sense that for each possible successful attack string **y**, our attack generation process can generate at least one string **z** such that z can be mapped to y. This means (1) the size **z** is less than or equal to size of **y**, and (2) **z** can be mapped to **y** such that if **y** is a successful attack string, than **z** is also a successful attack string. We have empirically evaluated this claim of coverage as will be described next section. Research is underway to formalize the steps of state machine construction described in this section in such a way that we can perform a more formal analysis of attack coverage.

It should be noted that browsers, particularly early versions of browsers, do not strictly follow W3C published standards. They tolerate syntax errors by rendering web pages that do not conform to markup language grammar. These exceptions can be accommodated in our approach as long as such exceptions are well defined. For example, we included generation of attack strings targeting IE7 in the attribute value context. Fortunately, modern browsers are converging by adhering more strictly to W3C standards. Finally as mentioned earlier, we need to extend the state machine to cover the entire JavaScript context.

## 4. EMPIRICAL EVALUATIONS

We conducted empirical evaluations of the effectiveness of our approach. First we applied this approach to one module of an open source project iTrust, a medical record management system with over 112K lines of Java/JSP code. Several XSS vulnerabilities were found. These vulnerabilities all stem from using html escaping for JavaScript context. A representative example is shown in Figure 10 along with a successful attack string generated by our attack generator. In this case the *request.getParameter()* is an untrusted source. The developer used HTML body encoding in a context where JavaScript encoding is needed. We scanned iTrust source code using Fortify SCA, a leading commercial static analysis tool. These vulnerabilities were not reported by Fortify SCA.

```
<input onclick= "parent.location.href= 'getPatientID.jsp?
forward= <%=escapeHtml("" + (
request.getParameter("forward") )) %> ';" />
```
*Attack vector: '; attack(); //*

**Figure 10. Sanitization Flaw found in iTrust**

Second, we looked at the performance of our attack generation. The experiments were performed on an iMac computer with a 2.8 GHz Intel core i7 with 8GB RAM. A total of 497 different attack strings can be generated based on the state machine in Figure 9. Generating and evaluating all these attack strings for a unit case takes 8.1 seconds of total execution time. Currently, once a successful attack script founds, the testing process stops and remaining none-tested attack scripts would not be checked and so cannot say how many successful attack scripts exist.

We evaluated the coverage of our attack generation model. We randomly selecting 400 attack scripts from a well-known repository of successful XSS attacks reported by penetration testers, *xssed.com*. This repository is often used by security researchers to evaluate the effectiveness of defense mechanisms (e.g. [1][8]). Selected scripts represent one percent of scripts in the repository. Our comparisons show 334 (83.5%) attack scripts corresponds exactly to at least one of the attack strings generated by our approach.

Each of the remaining 66 attack scripts can map to at least one of the attack strings generated by our approach. We illustrate a representative example: *{">><script> attack(); </script>}* is found in xssed.com. Our attack generator can generate *{' ><script> attack(); </script>}*, and *{"><script> attack(); </script>}*. The attack script in xsssed.com is a super string of both attack strings generated by our state machine. Although information about target application is not included in xssed.com, our analysis suggests that one of the two attack stings generated by our approach should be successful against that target.

## 5. DISCUSSION

Our approach may be applied to test for other injection vulnerabilities, such as command injection against shell scripts. Injection vulnerability is the largest class of software security flaws. It is well known that input validation is the most important defense against software vulnerabilities[5]. Wherever white list based input validation[5] is used, the question of whether the correct validation function is used for a particular application context naturally arises. Unit testing, using a similar approach as outlined here, can provide such assurance.

There are also cases where more than one sanitization functions must be used. In such cases, in addition to the issue of whether the correct sanitization function is used, one must also consider whether sanitization functions are applied in the correct order. Take the following example from a web application described in[6], a tainted variable is sanitized using two standard encoders of *escapeHtmlDecimal* and *escapeJavascript* before used in final sink <%= %>:

*<% **htmlEsc** =escapeHtmlDecimal( **Tainted**); %>*

*<input onclick="Fn('<%= escapeJavascriptl(**htmlEsc**) %>'); " type='button' />*

In this case, the order of encoders is incorrect. If the tainted value contains the single quote {'} the *escapeHtmlDecimal* escapes it to {'}. After this, the *escapeJavaScript* does nothing to this string. So an attack string {*');attack(); //*}will be changed to {');attack(); //}. Most browsers have an internal decoding feature, known as implicit transduction [15], that un-escapes a string before evaluation a JavaScript program. In this case the attack string is changed back to **'); attack();**// which successfully can exploit the vulnerable web application. The correct sequence is to apply *escapeJavaScript* before *escapeHtmlDecimal*. Our approach can detect this vulnerability.

There are other types of XSS vulnerability not addressed by our approach. Consider a blogging web site where one wants to permit end users the use of HTML markup tags for formatting. HTML body encoding cannot be used for blog content because HTML body encoding would disable all HTML markup tags. There are heuristic filter functions, e.g.[10] that try to block unwanted JavaScript programs in HTML body context. Such filters have proven to be very difficult to verify as many patches have been issued. Much industry effort such as [10,16] have been focused on testing this type of vulnerability. However, this type of vulnerability affects a relatively small fraction of web application functions. This is evidenced by the that fact a well referenced industry best practice XSS prevention programming guide[9] exclusively discusses how to match encoders to web application contexts.

## 6. CONCLUSION

In summary, we propose a unit test based approach to detect a large class of common cross-site scripting vulnerability caused by applying a wrong output encoder for a given application context. Our contributions are highlighted below. First the unit test approach can be easily integrated into integrated development environments and the software development process to detect security vulnerabilities early in the development[17]. Most software developers would be able to detect and correct these problems without engaging security experts. This can save valuable security testing resources to focus on other types of vulnerabilities. Second our approach can be applied to all web applications as well as newer applications written in template languages. Third this approach has the potential to be applied to detect other types of injection attacks. Our evaluation shows that the approach can be applied efficiently to detect vulnerabilities in large open source applications. Empirical study suggests our approach has good coverage of possible attacks. Current work is under way to more formally analyze attack coverage.

## 7. ACKNOWLEDGMENT

This research is support in part by NSF grants: 1129190, 1318854, 1318323.

## 8. REFRENCES